\documentclass[
 aps,
 pra,
 reprint,
 amsmath,
 amssymb,
 floatfix,
 superscriptaddress,
 longbibliography
]{revtex4-2}

\usepackage{bm}
\usepackage{booktabs}
\usepackage{braket}
\usepackage{graphicx}
\usepackage{placeins}
\usepackage[hidelinks]{hyperref}

\begin{document}

\title{Remote Flux Refocuses Nonadiabatic Excursions in Compact-State Quantum Transfer}

\author{X. S. Meng}
\affiliation{College of Physics and Materials Science, Tianjin Normal University, Tianjin 300387, China}
\author{Y. S. Liu}
\affiliation{College of Physics and Materials Science, Tianjin Normal University, Tianjin 300387, China}
\author{X. Z. Zhang}
\email{zhangxz@tjnu.edu.cn}
\affiliation{College of Physics and Materials Science, Tianjin Normal University, Tianjin 300387, China}
\affiliation{Interdisciplinary Center, Tianjin Normal University, Tianjin 300387, China}

\begin{abstract}
Compact states embedded in a propagating band can carry quantum information, but finite-time transfer can populate modes outside their subspace.
We show that a remote flux can refocus this amplitude without changing the transfer states or the motion-induced coupling out of their subspace.
In a resonator ring, this separation is exact because both endpoints of the flux-bearing bond are nodes of every compact transfer state.
For the same pulse, zero and refocusing flux both produce substantial excursions, but only the latter yields near-complete logical transfer.
Interference between matrix amplitudes associated with different windings suppresses the endpoint error.
Under matched control bounds, numerical optimization yields higher worst-input fidelity at the refocusing flux, providing a route to accurate transfer through coherent return.
\end{abstract}

\maketitle
\raggedbottom

\section{Introduction}

An excitation may remain localized even when its energy lies among propagating modes if destructive interference cancels every outgoing amplitude \cite{vonNeumann1929,Friedrich1985,Hsu2016,Koshelev2019}. In lattice systems, the same mechanism can produce compact states with exact nodes rather than exponentially small tails \cite{Flach2014,Leykam2018,Vicencio2015,Mukherjee2015}. Quantum communication relies on transfer between distant nodes \cite{Cirac1997,Bose2003,Christandl2004}. Compact-state protocols accomplish this through modulated couplings \cite{PRL2019CompactTransfer}. Because their energies are embedded in the propagating band, however, these states are not protected by a spectral gap from nearby modes. When the corresponding projector changes in finite time, amplitude is launched into those modes, and the endpoint error depends not only on how much amplitude leaves the encoded subspace but also on how it propagates and returns.

Adiabatic protocols reduce nonadiabatic departure by slowing the motion of a dark state \cite{Bergmann1998,Vitanov2017}, while counterdiabatic protocols cancel the transitions generated by that motion \cite{Berry2009}. Optimal control instead shapes the entire finite-time evolution and need not minimize intermediate occupation outside the target subspace \cite{Brif2010}. Interference between leakage amplitudes has also been used to suppress final leakage in superconducting quantum gates \cite{Rol2019LeakageInterference}. In a closely related coupled-resonator waveguide, Guo et al. demonstrated nonadiabatic transfer through bound states in the continuum (BICs), with intermediate occupation of neighboring scattering states and return to the BIC subspace at the endpoint \cite{Guo2026BICTransfer}. An embedded bound state can also be captured or reshaped without a protecting spectral gap \cite{Chang2026EmbeddedState}.

Phase-controlled interference has been used in non-Hermitian ring scatterers~\cite{Li2015FluxScattering,Jin2016ReciprocalScattering,Zhang2026ParityFlux}, directional amplification~\cite{Li2017DirectionalAmplification}, and flux-induced confinement~\cite{Zhang2019DynamicalConfinement}. Here we ask whether a remote phase can control return in a Hermitian compact-state transfer protocol while leaving the instantaneous transfer subspace, its energies, and its motion-induced coupling unchanged.

An exact sufficient condition gives this separation without specifying a lattice. Choose a moving transfer subspace independent of the remote-control parameter and preserved by the reference Hamiltonian, and require the added control to annihilate that subspace. Its motion then determines the coupling to the complement, while the remote parameter changes propagation there. Eliminating the complementary amplitudes yields a memory equation in which the parameter enters only through propagation between departure and return. This provides control of the return without directly changing the motion-induced coupling.

A finite resonator ring with three compact transfer states realizes this separation. A Peierls phase is placed on a remote bond whose endpoints are common nodes of the three states. For the same pulse, zero and refocusing flux both produce substantial excursions, but only the latter recovers nearly complete logical transfer. Resolving the returned amplitudes by their winding around the ring identifies interference among several matrix-valued sectors as the source of this recovery. Optimization within a common bounded control family shows that the interference can improve worst-input fidelity. Refocusing also occurs in a second ring, whose optimized protocol exhibits a different preferred flux and winding correlations.

The remainder of this paper is organized as follows. Section~II develops the exact dynamics under complement-only control. Section~III constructs the compact-state ring and identifies the Fourier components with winding sectors. Section~IV presents a fixed-pulse comparison that isolates the effect of the flux on complementary propagation and return. Section~V resolves the winding-dependent return error. Section~VI examines performance within bounded control resources, flux tolerance, and a changed circumference. Section~VII discusses the physical distinction from leakage-suppression strategies and the scope of the design condition, and Sec.~VIII concludes the article.

\section{Exact dynamics under complement-only control}

\subsection{A moving encoded subspace}

Consider a finite-dimensional Hilbert space, a differentiable Hermitian Hamiltonian
\begin{equation}
H_\chi(t)=H_0(t)+V_\chi(t)
\label{eq:general_hamiltonian}
\end{equation}
and a differentiable projector $P(t)$ of fixed rank $r$. Both $H_0(t)$ and $V_\chi(t)$ are Hermitian, and we assume

\begin{equation}
\begin{aligned}
\partial_\chi P(t)&=0,
& [H_0(t),P(t)]&=0,\\
V_\chi(t)P(t)&=0,
& P(t)V_\chi(t)&=0.
\end{aligned}
\label{eq:general_condition}
\end{equation}

\begin{figure*}[t]
\includegraphics[width=\textwidth]{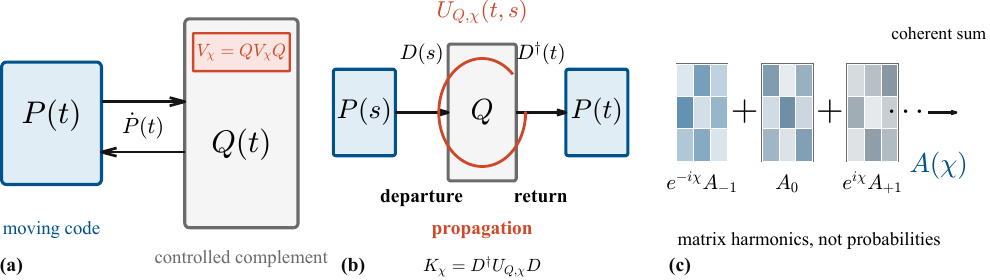}
\caption{Control confined to the complement. (a) Motion of $P(t)$ couples the encoded subspace and $Q(t)$ in both directions, while $V_\chi=QV_\chi Q$ acts only inside $Q$. (b) Exact elimination of $Q$ separates departure by $D(s)$, propagation by $U_{Q,\chi}(t,s)$, and return by $D^\dagger(t)$. (c) Periodic dependence on $\chi$ resolves the endpoint amplitude block into coherently summed matrix Fourier components. Their index remains a harmonic label until the ring construction assigns it a winding meaning.}
\label{fig:general_theory}
\end{figure*}

Because $V_\chi(t)$ is Hermitian and $P(t)=P^\dagger(t)$, the last two relations are Hermitian conjugates of one another; we display both to make the block structure explicit. The first relation means that the encoded subspace itself, not only its dimension, is independent of the control parameter. With $Q(t)=I-P(t)$, the last two relations are equivalent to
\begin{equation}
V_\chi(t)=Q(t)V_\chi(t)Q(t).
\label{eq:complement_only}
\end{equation}
Here the code is simply the chosen $r$-dimensional encoded subspace; no error-correcting structure is assumed. The control has no matrix element connecting this subspace either to itself or to its complement. Figure~\ref{fig:general_theory}(a) summarizes this block structure.

Let the columns of $B(t)$ and $C(t)$ form orthonormal frames for $P(t)$ and $Q(t)$, respectively. Expanding $|\psi(t)\rangle=B(t)a(t)+C(t)b(t)$, define
\begin{align}
\mathcal E(t)&=B^\dagger H_0B-iB^\dagger\dot B,
\label{eq:general_E}\\
\mathcal H_{Q,\chi}(t)&=C^\dagger H_\chi C-iC^\dagger\dot C,
\label{eq:general_HQ}\\
D(t)&=C^\dagger(t)\dot B(t).
\label{eq:general_D}
\end{align}
All operators in Eqs.~(\ref{eq:general_E})--(\ref{eq:general_D}) are written in the moving frames; $\mathcal E(t)$ is the effective generator inside the moving encoded subspace. Because $P(t)$ and $Q(t)$ are independent of $\chi$, we choose orthonormal frames $B(t)$ and $C(t)$ that are also independent of $\chi$. With this choice, Eq.~(\ref{eq:general_condition}) makes both $\mathcal E(t)$ and $D(t)$ independent of $\chi$, whereas $\mathcal H_{Q,\chi}(t)$ retains the full control dependence. No parallel-transport condition is required for this separation. Other $\chi$-independent frame choices give the same physical separation through the covariance derived in Appendix~\ref{app:moving_projector}.

\subsection{Exact departure, propagation, and return}

Projection of the Schr\"odinger equation onto the two moving frames gives
\begin{align}
\dot a(t)&=-i\mathcal E(t)a(t)+D^\dagger(t)b(t),
\label{eq:a_projected}\\
\dot b(t)&=-D(t)a(t)-i\mathcal H_{Q,\chi}(t)b(t).
\label{eq:b_projected}
\end{align}
The coupling $D(t)$ arises from motion of the encoded subspace. It transfers amplitude from $P(t)$ into $Q(t)$, while $D^\dagger(t)$ couples it back. Both are fixed when $\chi$ is changed.

Let the complementary propagator satisfy
\begin{align}
i\partial_tU_{Q,\chi}(t,s)&=\mathcal H_{Q,\chi}(t)U_{Q,\chi}(t,s),
\nonumber\\
U_{Q,\chi}(s,s)&=I_Q.
\label{eq:UQ_definition}
\end{align}
For an initial state in the encoded subspace, $b(0)=0$, Eq.~(\ref{eq:b_projected}) has the exact solution
\begin{equation}
b(t)=-\int_0^t U_{Q,\chi}(t,s)D(s)a(s)\,ds.
\label{eq:duhamel}
\end{equation}
Substitution into Eq.~(\ref{eq:a_projected}) yields
\begin{equation}
\dot a(t)=-i\mathcal E(t)a(t)-\int_0^tK_\chi(t,s)a(s)\,ds,
\label{eq:memory_equation}
\end{equation}
where
\begin{equation}
K_\chi(t,s)=D^\dagger(t)U_{Q,\chi}(t,s)D(s).
\label{eq:memory_kernel}
\end{equation}
Equations~(\ref{eq:memory_equation}) and (\ref{eq:memory_kernel}) are exact and contain no adiabatic, perturbative, Markov, or few-mode approximation. The parameter $\chi$ leaves the local generator and the couplings between the two subspaces unchanged, but modifies propagation in the complement. The full encoded dynamics can therefore depend on $\chi$ through the memory kernel. Its three factors are represented in Fig.~\ref{fig:general_theory}(b).

For fixed input and output isometries $\mathsf S,\mathsf R:\mathbb C^r\rightarrow\mathcal H$, define the endpoint amplitude block

\begin{equation}
A(\chi)=\mathsf R^\dagger U_\chi(T,0)\mathsf S.
\label{eq:A_general}
\end{equation}

For the smooth $2\pi$-periodic parameter dependence considered here, the propagator, kernel, and endpoint block have convergent Fourier series,
\begin{align}
U_{Q,\chi}(t,s)&=\sum_{\ell\in\mathbb Z}e^{i\ell\chi}U_{Q,\ell}(t,s),
\label{eq:UQ_fourier}\\
K_\chi(t,s)&=\sum_{\ell\in\mathbb Z}e^{i\ell\chi}K_\ell(t,s),
\label{eq:kernel_fourier}\\
K_\ell(t,s)&=D^\dagger(t)U_{Q,\ell}(t,s)D(s).
\end{align}
The endpoint block consequently has the form

\begin{equation}
A(\chi)=\sum_{\ell\in\mathbb Z}e^{i\ell\chi}A_\ell.
\label{eq:A_general_fourier}
\end{equation}

For a multidimensional encoded subspace, the Fourier coefficients are matrices rather than probabilities. In the general construction their index labels harmonics of the control parameter. The ring realization below gives this index a physical interpretation as the net number of crossings of a chosen cut.

\section{Compact-state ring realization}

Throughout, $\hbar=1$ and $J$ sets the energy scale. We use the coupled-resonator and emitter-register architecture considered in Ref.~\cite{Guo2026BICTransfer} and introduce a remote bond with independently tunable magnitude and phase. Energies are measured relative to the common cavity frequency, so the cavity onsite energy is zero and $\Omega_\alpha$ denotes an emitter detuning from that reference. This architecture provides a representative waveguide-QED setting \cite{Sheremet2023}. Consider $N_c$ single-mode resonators with nearest-neighbor hopping $J>0$ and site labels understood modulo $N_c$. Two registers, each containing $N_a$ distinguishable two-level emitters, are attached at cavities $n_s$ and $n_r$. Let $|n\rangle$ denote one photon at cavity $n$, and let $|s,\alpha\rangle$ and $|r,\alpha\rangle$ denote an excitation of sender or receiver emitter $\alpha=1,\ldots,N_a$. The flux lies on a bond $n_\phi\leftrightarrow n_\phi+1$ along the long arc between the registers.

\begin{figure*}[t]
\includegraphics[width=\textwidth]{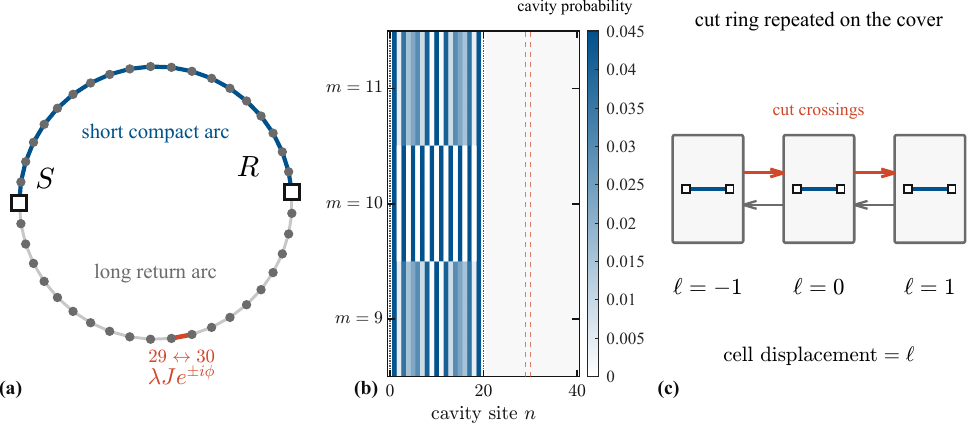}
\caption{The phase-bearing bond lies entirely outside the compact-state support, and cutting that bond converts the flux harmonics into winding sectors. (a) For $N_c=41$, $N_a=3$, $n_s=0$, $n_r=20$, and $n_\phi=29$, the sender $S$ and receiver $R$ terminate the short arc supporting the compact states, while the phase-bearing bond lies on the long return arc. (b) Cavity probabilities of the three modes at $g_s=g_r=0.4J$. The cavity amplitudes vanish at the two attachment sites $n=0,20$ and throughout the long arc. (c) Cutting the remote bond and repeating the opened graph produces the cyclic cover. Each cut crossing changes the cell coordinate, and a displacement $\ell$ represents net winding $\ell$.}
\label{fig:ring_realization}
\end{figure*}

We separate this distinguished bond from the remaining terms so that the ring model has the form of Eq.~(\ref{eq:general_hamiltonian}): $H_0(t)$ omits the bond and $V_{\phi,\lambda}$ supplies it. The parameter $\lambda\in[0,1]$ controls the bond magnitude. The gauge-invariant phase accumulated around the ring is $\phi$; in the gauge used here, that phase is placed entirely on the distinguished bond \cite{Peierls1933,Aharonov1959}. At $\lambda=0$, the distinguished bond is absent and the resonator graph is an open chain. At $\lambda=1$, it is a closed ring. The open-link case is therefore a connectivity control for the present model, whereas comparing zero and nonzero flux at $\lambda=1$ holds the graph fixed. The numerical realization used below has $N_c=41$, $N_a=3$, $n_s=0$, $n_r=d=20$, and $n_\phi=29$, as shown in Fig.~\ref{fig:ring_realization}(a). Its single-excitation Hilbert space has dimension $47$. Explicitly,

\begin{align}
H(t;\phi,\lambda)={}&H_0(t)+V_{\phi,\lambda},
\label{eq:hamiltonian_split}\\
H_0(t)={}&-J\!\sum_{\substack{n=0\\n\ne n_\phi}}^{N_c-1}
\left(|n\rangle\langle n+1|+\mathrm{H.c.}\right)
\nonumber\\
&+\sum_{\alpha=1}^{N_a}\Omega_\alpha
\left(|s,\alpha\rangle\langle s,\alpha|
+|r,\alpha\rangle\langle r,\alpha|\right)
\nonumber\\
&+\sum_{\alpha=1}^{N_a}\left[g_s(t)|n_s\rangle\langle s,\alpha|
+\mathrm{H.c.}\right]
\nonumber\\
&+\sum_{\alpha=1}^{N_a}\left[g_r(t)|n_r\rangle\langle r,\alpha|
+\mathrm{H.c.}\right],
\label{eq:hamiltonian}\\
V_{\phi,\lambda}={}&-\lambda J e^{i\phi}
|n_\phi\rangle\langle n_\phi+1|\nonumber\\
&-\lambda J e^{-i\phi}
|n_\phi+1\rangle\langle n_\phi|.
\label{eq:ring_control}
\end{align}

The three emitter frequencies are commensurate with the short arc,
\begin{align}
k_\alpha&=\frac{m_\alpha\pi}{d},
&\Omega_\alpha&=-2J\cos k_\alpha,
\nonumber\\
(m_1,m_2,m_3)&=(11,10,9).&&
\label{eq:resonance}
\end{align}
An arbitrary encoded input is $|\psi_s\rangle=\sum_\alpha c_\alpha|s,\alpha\rangle$, with $\sum_\alpha|c_\alpha|^2=1$. The ring is finite, so $Q(t)$ is the finite-dimensional orthogonal complement of the three compact states. It contains the discretized descendants of the propagating band; no continuum-limit statement is used below.

\subsection{Exact compact transfer states}

To keep the compact basis well defined throughout the pulse, we take $g_s(t),g_r(t)\ge0$ and impose
\begin{equation}
g_s^2(t)+g_r^2(t)>0\qquad (0\le t\le T).
\label{eq:nonvanishing_couplings}
\end{equation}
For each channel, define $\eta_\alpha=(-1)^{m_\alpha+1}$. The normalized instantaneous eigenstate
\begin{align}
|B_\alpha(t)\rangle=\frac{1}{\mathcal N_\alpha(t)}\biggl\{&J\sin k_\alpha
\left[g_r(t)|s,\alpha\rangle+\eta_\alpha g_s(t)|r,\alpha\rangle\right]
\nonumber\\
&+g_s(t)g_r(t)\sum_{n=1}^{d-1}\sin(k_\alpha n)|n\rangle\biggr\}
\label{eq:compact_state}
\end{align}
has energy $\Omega_\alpha$, where
\begin{equation}
\mathcal N_\alpha^2(t)=J^2\sin^2k_\alpha\,[g_s^2(t)+g_r^2(t)]
+\frac{d}{2}g_s^2(t)g_r^2(t).
\label{eq:normalization}
\end{equation}
The photonic profile is a standing wave on the short arc. The emitter amplitudes cancel hopping through its endpoints, and discrete sine orthogonality makes the three states mutually orthogonal. Their cavity probabilities over the full ring appear in Fig.~\ref{fig:ring_realization}(b). Direct substitution gives
\begin{equation}
H(t;\phi,\lambda)|B_\alpha(t)\rangle
=\Omega_\alpha|B_\alpha(t)\rangle
\label{eq:compact_eigenvalue}
\end{equation}
for every $\phi$ and $\lambda$. Appendix~\ref{app:compact} derives the endpoint cancellations, normalization, orthogonality, and regular endpoint limits. Each state has exactly zero amplitude on the long arc, including both endpoints of the phase-bearing bond.
Their common zeros make the remote bond an exact location for a control that is invisible to the compact subspace but active in its complement.

With $g_s(0)=g_r(T)=0$ and $g_r(0)=g_s(T)=g_{\max}>0$, the compact basis connects the physical registers according to

\begin{equation}
|B_\alpha(0)\rangle=|s,\alpha\rangle,
\qquad
|B_\alpha(T)\rangle=\eta_\alpha|r,\alpha\rangle.
\label{eq:endpoint_mapping_main}
\end{equation}

Collect the compact states into
\begin{equation}
B(t)=\bigl(|B_1(t)\rangle,|B_2(t)\rangle,|B_3(t)\rangle\bigr),
\qquad P(t)=B(t)B^\dagger(t).
\label{eq:compact_projector}
\end{equation}
For the remote-bond operator, the common nodes imply
\begin{equation}
V_{\phi,\lambda}P(t)=P(t)V_{\phi,\lambda}=0.
\label{eq:two_sided}
\end{equation}
At $\lambda=0$, Eq.~(\ref{eq:compact_eigenvalue}) also shows that $H_0(t)$ preserves the range of $P(t)$. Thus Eqs.~(\ref{eq:hamiltonian_split}), (\ref{eq:ring_control}), and (\ref{eq:two_sided}) realize every condition in Eq.~(\ref{eq:general_condition}).
The explicit real basis is also parallel, $B^\dagger\dot B=0$, and
\begin{equation}
\mathcal E=E_D=\operatorname{diag}(\Omega_1,\Omega_2,\Omega_3).
\label{eq:encoded_energy}
\end{equation}
We may also choose a parallel complementary frame, $C^\dagger\dot C=0$. The general equations of Sec.~II then reduce to Eqs.~(\ref{eq:a_projected}) and (\ref{eq:b_projected}) with $\mathcal E=E_D$ and $\mathcal H_{Q,\chi}=C^\dagger HC$. Appendix~\ref{app:moving_projector} gives the projector identities and frame covariance.

\subsection{Remote flux and winding sectors}

For the winding analysis, we set $\lambda=1$ and vary only the flux $\phi$. Cutting the phase-bearing bond and repeating the opened graph generates an infinite cyclic cover. Translation by one cover cell records one net crossing of the cut. A Bloch transform over the cell coordinate identifies the Bloch phase with the physical flux, so the harmonic index introduced in Sec.~II acquires the concrete interpretation of a net winding number. With $\mathsf S$ and $\mathsf R$ denoting the physical sender and receiver embeddings, the endpoint map is

\begin{equation}
A(\phi)\equiv \mathsf R^\dagger U(T,0;\phi,\lambda=1)\mathsf S
=\sum_{\ell\in\mathbb Z}e^{i\ell\phi}A_\ell,
\label{eq:A_fourier}
\end{equation}
where $A_\ell$ is the sender-to-receiver block associated with net winding $\ell$. The corresponding cell displacement on the cover is drawn in Fig.~\ref{fig:ring_realization}(c). Individual sector matrices depend on the fixed cut and endpoint convention, whereas their coherent reconstruction and every logical fidelity are physical. Appendix~\ref{app:cover} derives the cyclic-cover transform and fixes this convention.

Holding the pulse fixed keeps the compact basis and the coupling $D(t)$ unchanged, so varying the flux isolates the effect of complementary propagation on the endpoint transfer.
\FloatBarrier
\section{Flux-controlled return at a fixed pulse}
\label{sec:fixed_pulse}

\begin{figure*}[t]
\includegraphics[width=\textwidth]{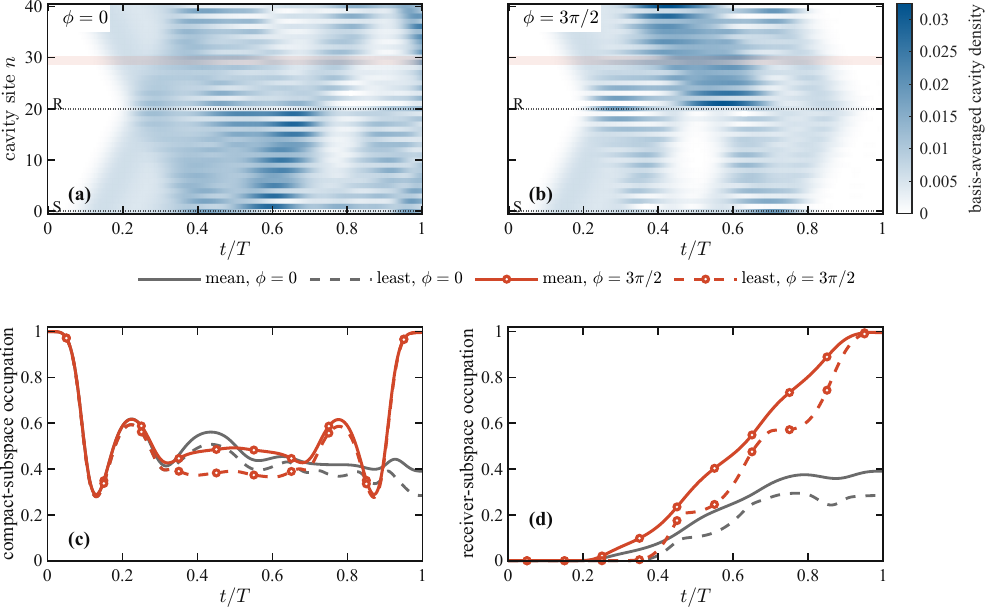}
\caption{Flux-controlled return for one fixed pulse. (a,b) Basis-averaged cavity densities at zero flux and $\phi_\star=3\pi/2$ on a common color scale; the pale band marks the cut. (c,d) Mean and minimum occupations of the compact and receiver subspaces, respectively, defined in Eq.~(\ref{eq:subspace_occupations}). The closed ring has $N_c=41$, $N_a=3$, $d=20$, $\lambda=1$, and $JT=75$. Only the flux changes between the two evolutions.}
\label{fig:same_control}
\end{figure*}

\subsection{Fixed protocol and observables}

Let $x_\star$ denote the endpoint-coupling pulse obtained from the constrained numerical optimization at
\begin{equation}
\phi_\star=\frac{3\pi}{2},\qquad JT=75.
\end{equation}
We refer to $\phi_\star$ as the refocusing flux.

The ideal endpoint map is
\begin{equation}
A_{\rm id}=\operatorname{diag}\left(\eta_1e^{-i\Omega_1T},
\eta_2e^{-i\Omega_2T},\eta_3e^{-i\Omega_3T}\right).
\end{equation}
The known ideal endpoint phases and parity signs are removed through $A_{\rm id}^{-1}$, after which two receiver-local relative phases remain adjustable. The resulting diagonal calibration
\begin{equation}
Z(\zeta_2,\zeta_3)=\operatorname{diag}(1,e^{i\zeta_2},e^{i\zeta_3})A_{\rm id}^{-1},
\label{eq:Z_calibration}
\end{equation}
cannot mix the three logical channels. The same constrained optimization determines receiver phases $(\zeta_{2,\star},\zeta_{3,\star})$; we denote the resulting diagonal calibration by

\begin{equation}
Z_\star=Z(\zeta_{2,\star},\zeta_{3,\star}).
\label{eq:Z_star}
\end{equation}

The pulse $x_\star$ and calibration $Z_\star$ are then held fixed when the flux is changed to zero and throughout the detuning scan. We propagate this same protocol at $\phi=0$ and $\phi=\phi_\star$. The pulse, duration, moving projector, compact energies, and motion-induced coupling $D(t)$ are identical in the two evolutions; only the complementary propagator changes. A full $U(3)$ correction is excluded because it could undo transfer-induced channel mixing rather than merely calibrate known receiver-local phases. For the transfer-amplitude block $A$, define
\begin{equation}
F_{\rm logic}(Z,A)=\min_{\|c\|=1}|c^\dagger ZAc|^2.
\label{eq:Flogic}
\end{equation}
This worst-input target overlap penalizes incomplete receiver population, unequal attenuation, phase error, and coherent channel mixing. It is not a postselected fidelity or the process fidelity of a trace-preserving quantum channel.
For the fixed pulse $x_\star$, define

\begin{equation}
F_{\rm logic}^{\rm fix}(\phi)
=F_{\rm logic}\!\left[Z_\star,A(x_\star;\phi)\right].
\label{eq:Flogic_fixed}
\end{equation}

For $X(t;\phi)=U(t,0;\phi)\mathsf S$, where $\mathsf S$ embeds the three sender amplitudes, the basis-averaged cavity density shown in Figs.~\ref{fig:same_control}(a) and \ref{fig:same_control}(b) is

\begin{equation}
\bar p_n(t;\phi)
=
\frac{1}{3}
\sum_{\alpha=1}^{3}
\left|
\langle n|
U(t,0;\phi)
|s,\alpha\rangle
\right|^2.
\label{eq:basis_averaged_cavity_density}
\end{equation}

The compact- and receiver-subspace occupations are
\begin{align}
\overline p_P(t;\phi)&=\frac{1}{3}\operatorname{Tr}\left[X^\dagger P(t)X\right],
\nonumber\\
p_P^{\min}(t;\phi)&=\lambda_{\min}\left[X^\dagger P(t)X\right],
\nonumber\\
\overline p_R(t;\phi)&=\frac{1}{3}\operatorname{Tr}\left[X^\dagger P_RX\right],
\nonumber\\
p_R^{\min}(t;\phi)&=\lambda_{\min}\left[X^\dagger P_RX\right],
\label{eq:subspace_occupations}
\end{align}
where $P_R=\sum_\alpha|r,\alpha\rangle\langle r,\alpha|$. The $P$ quantities measure occupation of the moving compact subspace and therefore quantify the actual excursion, which need not be flux independent even though the geometric coupling $D(t)$ is. The $P_R$ quantities measure returned population without assuming that the receiver phases and channel structure are correct.

At the endpoint, the least receiver population is

\begin{equation}
P_R^{\min}(\phi)=p_R^{\min}(T;\phi)
=\lambda_{\min}\!\left[A^\dagger(\phi)A(\phi)\right].
\label{eq:least_receiver_population}
\end{equation}

Because $A_{\rm id}$ and the additional receiver phase matrix are diagonal and unitary, $Z$ is unitary. Therefore, $|c^\dagger ZAc|^2\le c^\dagger A^\dagger Ac$ for every normalized $c$, and hence

\begin{equation}
F_{\rm logic}(Z,A)\le P_R^{\min}.
\label{eq:receiver_population_bound}
\end{equation}

A small least receiver population therefore cannot be repaired by changing only receiver-local phases. Appendix~\ref{app:fidelity} gives the complete proof.

\subsection{Substantial departure and different return}

Figures~\ref{fig:same_control}(a) and \ref{fig:same_control}(b) compare the full 47-dimensional evolution at zero and refocusing flux. In both cases, amplitude leaves the compact arc and reaches the long return arc. The mean and least-retained compact occupations fall substantially below unity at both fluxes [Fig.~\ref{fig:same_control}(c)]. High endpoint fidelity therefore coexists with substantial nonadiabatic departure from the compact subspace.

The difference emerges during return. At zero flux, the receiver population remains incomplete [Fig.~\ref{fig:same_control}(d)], and the bound in Eq.~(\ref{eq:receiver_population_bound}) proves that a different diagonal phase calibration cannot repair this deficit. At $\phi_\star$, the same pulse transfers every encoded input state to the receiver with high fidelity. With the receiver calibration fixed at the refocusing point, the endpoint fidelities are

\begin{equation}
F_{\rm logic}^{\rm fix}(0)=0.282,
\qquad
F_{\rm logic}^{\rm fix}(\phi_\star)=0.995.
\label{eq:fixed_control_fidelities}
\end{equation}
Since all other ingredients are identical, the endpoint contrast originates in the complementary propagator. Its winding-resolved matrix structure, rather than the total occupation outside the compact subspace, controls the phases and channel mixing of the returning amplitudes.

\FloatBarrier
\section{Winding-resolved return error}

\begin{figure*}[t]
\includegraphics[width=\textwidth]{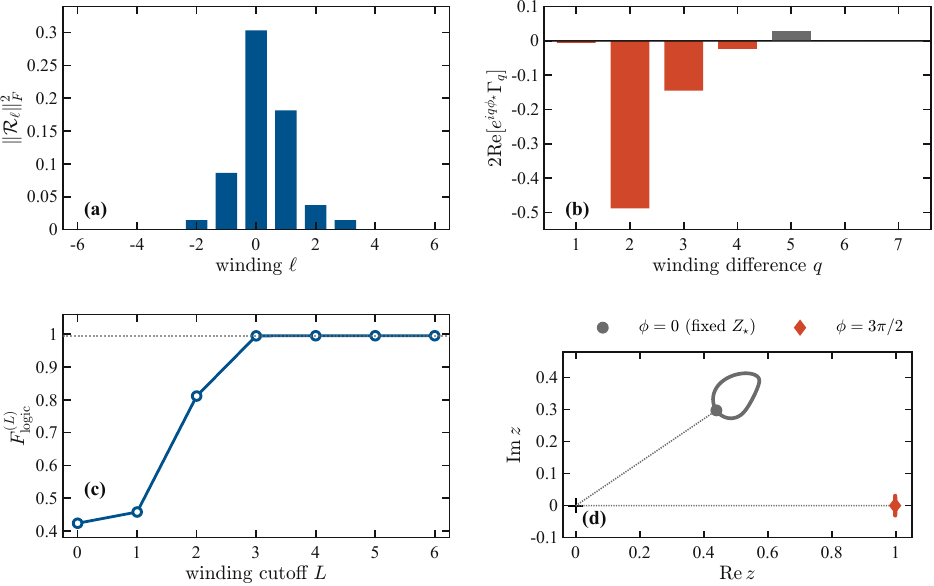}
\caption{Winding-resolved cancellation of the return error. The pulse and phase conventions are those of the $N_c=41$, $JT=75$ refocusing protocol. (a) Sector powers $\|\mathcal R_\ell\|_F^2$. (b) Signed contributions $2\operatorname{Re}[e^{iq\phi_\star}\Gamma_q]$; negative values reduce the endpoint error. (c) Worst-input fidelity $F_{\rm logic}^{(L)}=F_{\rm logic}[Z_\star,A^{(L)}(\phi_\star)]$ reconstructed from $|\ell|\le L$. Several shells are required before the coherent matrix sum converges. (d) Numerical ranges of $e^{-i\theta_\star}Z_\star A(\phi)$ at zero and refocusing flux. The thin dotted origin-to-boundary segments give the distances entering Eq.~(\ref{eq:numerical_range}). The receiver calibration $Z_\star$ and common phase $\theta_\star$ are fixed throughout.}
\label{fig:winding}
\end{figure*}

\subsection{Fixed-calibration return matrix}

The winding analysis uses one physical receiver calibration for all fluxes. After determining $Z_\star$ at $\phi_\star$, we choose the physically irrelevant global phase that minimizes the Frobenius distance of the nominal calibrated map from the identity,

\begin{equation}
\theta_\star=\arg\operatorname{Tr}\left[Z_\star A(\phi_\star)\right].
\label{eq:theta_star}
\end{equation}

Appendix~\ref{app:fidelity} derives this choice. Both $Z_\star$ and $\theta_\star$ are then fixed for every flux. The return-error matrix is
\begin{equation}
\mathcal R(\phi)=I_3-e^{-i\theta_\star}Z_\star A(\phi)
=\sum_{\ell\in\mathbb Z}e^{i\ell\phi}\mathcal R_\ell,
\label{eq:return_error}
\end{equation}
with
\begin{equation}
\mathcal R_0=I_3-e^{-i\theta_\star}Z_\star A_0,
\qquad
\mathcal R_{\ell\ne0}=-e^{-i\theta_\star}Z_\star A_\ell.
\label{eq:return_sectors}
\end{equation}
The identity contributes only to $\mathcal R_0$. A flux-dependent recalibration would mix calibration-induced harmonics with the physical winding sectors.

\subsection{Cross-winding cancellation}

Parseval's identity gives the phase-independent reference
\begin{equation}
P_{\rm inc}=\sum_\ell\|\mathcal R_\ell\|_F^2
=\frac{1}{2\pi}\int_0^{2\pi}\|\mathcal R(\phi)\|_F^2\,d\phi.
\label{eq:Pinc}
\end{equation}
Thus $P_{\rm inc}$ is the phase-averaged squared return error, equal to the sum of the sector powers with all cross-winding terms omitted. This average does not describe a separate physical decoherence process.
Define the winding-difference correlations

\begin{equation}
\Gamma_q=\sum_{\ell\in\mathbb Z}
\operatorname{Tr}\left(\mathcal R_\ell^\dagger \mathcal R_{\ell+q}\right),
\qquad q\ge1.
\label{eq:Gammaq}
\end{equation}

The total return-error power is then
\begin{equation}
\|\mathcal R(\phi)\|_F^2=P_{\rm inc}
+2\sum_{q=1}^{\infty}\operatorname{Re}\left[e^{iq\phi}\Gamma_q\right].
\label{eq:error_harmonics}
\end{equation}
We denote the coherent cross-winding contribution by
\begin{equation}
P_{\rm cross}(\phi)=2\sum_{q=1}^{\infty}
\operatorname{Re}\left[e^{iq\phi}\Gamma_q\right],
\label{eq:Pcross}
\end{equation}
so that $\|\mathcal R(\phi)\|_F^2=P_{\rm inc}+P_{\rm cross}(\phi)$. With these conventions, the $N_c=41$, $JT=75$ protocol gives
\begin{align}
P_{\rm inc}&=0.6372,
\nonumber\\
P_{\rm cross}(\phi_\star)&=-0.6353,
\nonumber\\
\|\mathcal R(\phi_\star)\|_F^2&=1.94\times10^{-3}.
\label{eq:cancellation_numbers}
\end{align}
The ratio
\[
-\frac{P_{\rm cross}(\phi_\star)}{P_{\rm inc}}=0.997
\]
means that cross-winding interference cancels $99.7\%$ of the incoherent sum $P_{\rm inc}$. Several sectors with positive and negative winding numbers carry appreciable power [Fig.~\ref{fig:winding}(a)], and their negative cross terms nearly cancel the summed sector powers [Fig.~\ref{fig:winding}(b)]. This is cancellation of error power relative to a fixed reference map, not a population-loss fraction. Logical fidelity is insensitive to a common phase and is evaluated separately below.

\subsection{Worst-input fidelity and numerical range}

For a fixed encoded input $c$, the calibrated target amplitude is $c^\dagger ZAc$. The set of these complex amplitudes over all normalized inputs is the numerical range of $ZA$. For a matrix $M$, its numerical range is \cite{Gustafson1997,HornJohnson2013}
\begin{equation}
W(M)=\{c^\dagger Mc:\|c\|=1\}.
\end{equation}
The worst-input fidelity can be written as
\begin{equation}
F_{\rm logic}(Z,A)=d^2[0,W(ZA)],
\label{eq:numerical_range}
\end{equation}
where $d[0,W]$ is the Euclidean distance from the origin to $W$. Appendix~\ref{app:fidelity} distinguishes this logical quantity from the smallest receiver population. For a symmetric winding truncation,
\begin{equation}
A^{(L)}(\phi_\star)=\sum_{|\ell|\le L}e^{i\ell\phi_\star}A_\ell.
\label{eq:symmetric_winding}
\end{equation}
For every truncation we retain the receiver calibration determined from the full refocusing protocol and define

\begin{equation}
F_{\rm logic}^{(L)}
\equiv
F_{\rm logic}\!\left[
Z_\star,
A^{(L)}(\phi_\star)
\right].
\label{eq:truncated_fidelity}
\end{equation}

The truncated matrices are partial sums, not separately implemented transfer protocols. With the full-protocol calibration, $F_{\rm logic}^{(L)}$ approaches its final value only after several winding shells are included [Fig.~\ref{fig:winding}(c)]. The reconstruction is a coherent matrix sum rather than an additive probability sum. At the refocusing flux, the numerical range moves away from the origin [Fig.~\ref{fig:winding}(d)], expressing the recovery of the target overlap across all encoded inputs.

Refocusing thus recovers the receiver map with high fidelity for encoded superpositions, rather than only transferring population from one initial state. The same mechanism can improve transfer under bounded pulse resources.

\FloatBarrier
\section{Performance within bounded control resources}

\subsection{Matched-resource comparison}

\begin{figure*}[t]
\includegraphics[width=\textwidth]{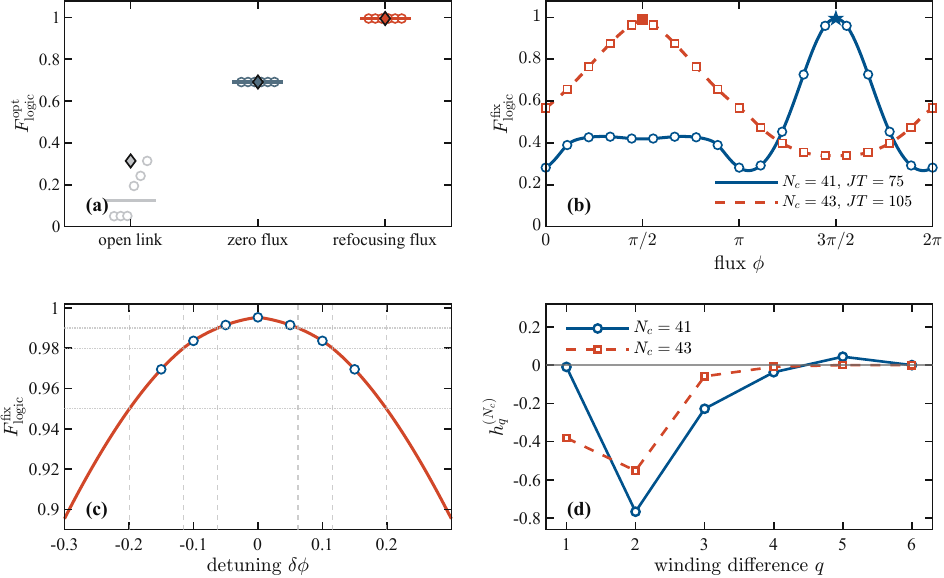}
\caption{Remote-flux refocusing under constrained optimization, finite flux error, and a changed return geometry. Results in (a) are separately optimized; (b)--(d) keep each pulse and receiver calibration fixed at its respective refocusing point. (a) Six runs per intervention for $N_c=41$, $JT=75$, under the same ten-parameter control family, coefficient and amplitude bounds, endpoint conditions, and receiver-phase freedom. The zero- and refocusing-flux runs use paired initial controls. Small open circles denote individual runs, horizontal bars their medians, and filled diamonds the highest values quoted in Eq.~(\ref{eq:optimized_cases}). (b) Full-period fixed-pulse responses for $N_c=41$, $JT=75$ (solid blue), and $N_c=43$, $JT=105$ (dashed orange). The two ring protocols use the distinct pulse-selection and calibration prescriptions specified in Sec.~\ref{sec:second_ring}. Lines are winding reconstructions; open symbols are separate direct propagations, and filled symbols mark the refocusing points. (c) Fixed-pulse detuning of the $N_c=41$ protocol about $\phi_\star^{(41)}=3\pi/2$. Lines and open symbols have the same meaning as in (b), and the vertical guides mark the numerical $0.99$, $0.98$, and $0.95$ crossings obtained from direct propagation. (d) Normalized winding-difference contributions $h_q^{(N_c)}$ defined in Eq.~(\ref{eq:hq}). Negative $h_q^{(N_c)}$ values reduce the error at the corresponding refocusing flux. Both ring protocols support refocusing, with different phases and winding correlations.}
\label{fig:performance}
\end{figure*}

All three interventions are optimized at $JT=75$ within the same ten-parameter pulse family of Appendix~\ref{app:control}, with identical coefficient and amplitude bounds, endpoint conditions, frequency content of the pulse family, and receiver-phase freedom. Unlike the fixed-pulse comparison above, the pulse and receiver phases are optimized separately for each intervention. The superscript $\mathrm{opt}$ denotes the highest value found for each intervention in this constrained search:
\begin{equation}
F_{\rm logic}^{\rm opt}=
\begin{cases}
0.3137, & \lambda=0 \quad \text{(open link)},\\
0.6915, & \lambda=1,\ \phi=0,\\
0.9954, & \lambda=1,\ \phi=3\pi/2.
\end{cases}
\label{eq:optimized_cases}
\end{equation}
Opening the link removes one route and changes connectivity, so the first value is a baseline. The two closed-ring cases differ in the imposed flux but share the same admissible pulse family and receiver-phase freedom; their pulses are optimized separately. The comparison therefore measures the performance accessible under matched control constraints, whereas the fixed-pulse comparison of Sec.~\ref{sec:fixed_pulse} isolates the effect of changing the flux on one protocol. Across six paired optimizations, the refocusing-flux improvement over zero flux remains approximately $0.304$; the open-link baseline uses six separate starts. These values are not certified global maxima within the pulse family. Appendix~\ref{app:control} defines the common bounded pulse family and its endpoint and frequency constraints.

\subsection{Flux response and detuning}

For the $N_c=41$ protocol, Eq.~(\ref{eq:Flogic_fixed}) defines the flux response without retuning either the pulse or the receiver calibration. Direct propagation and the winding reconstruction agree to plotting resolution over the full period [Fig.~\ref{fig:performance}(b)]. For the fixed pulse and receiver calibration, the connected intervals around $\phi_\star^{(41)}$ with worst-input fidelities at least $0.99$, $0.98$, and $0.95$ have approximate half-widths
\begin{equation}
\begin{aligned}
\delta\phi_{0.99}&\simeq0.0625, &
\delta\phi_{0.98}&\simeq0.116,\\
\delta\phi_{0.95}&\simeq0.199,
\end{aligned}
\label{eq:flux_widths}
\end{equation}
respectively, in radians.
Quasistatic Gaussian flux noise damps the contribution with winding difference $q$ by $\exp(-q^2\sigma_\phi^2/2)$, as derived in Appendix~\ref{app:winding}. Interference between sectors farther apart in winding number is therefore more sensitive to flux noise. This law gives the noise-averaged squared Frobenius norm of the return error at fixed calibration, not the average worst-input fidelity.

\subsection{Changed return geometry}
\label{sec:second_ring}

The same mechanism survives a change in the return path. For $N_c=43$ at the same register separation $d=20$, the remote bond is $30\leftrightarrow31$, and the compact-state construction remains exact while the long arc changes. For this second example, $\Delta\Omega=2\pi/T$ replaces the spacing used for $N_c=41$, while the five harmonics and the amplitude and coefficient bounds are retained. Pulse selection uses a smooth approximation to the minimum receiver population. The logical fidelity is evaluated afterward with a diagonal phase calibration, rather than optimized jointly with the receiver phases as in the $N_c=41$ comparison. At $JT=105$, the pulses $x_\star$ and $x_0$ obtained at $\phi=\pi/2$ and $\phi=0$ give
\begin{equation}
\begin{aligned}
F_{\rm logic}[Z_\star,A(x_\star;\pi/2)]&\simeq0.990,\\
F_{\rm logic}[Z_0,A(x_0;0)]&\simeq0.812.
\end{aligned}
\label{eq:Nc43_values}
\end{equation}
For each protocol, the diagonal calibration is chosen to make the diagonal entries of its nominal transfer block real and nonnegative, up to an irrelevant common phase. The same calibration is then held fixed throughout the associated flux scan. Holding $x_\star$ and its nominal calibration $Z_\star$ fixed gives the second flux response in Fig.~\ref{fig:performance}(b). It peaks at $\phi_\star^{(43)}=\pi/2$, rather than at $\phi_\star^{(41)}=3\pi/2$. For each circumference, define the normalized winding-difference contribution

\begin{equation}
h_q^{(N_c)}=\frac{2\operatorname{Re}\left[
e^{iq\phi_\star^{(N_c)}}\Gamma_q^{(N_c)}\right]}
{P_{\rm inc}^{(N_c)}}.
\label{eq:hq}
\end{equation}

The zero-flux value on the fixed-pulse curve in Fig.~\ref{fig:performance}(b) is distinct from the separately optimized zero-flux value in Eq.~(\ref{eq:Nc43_values}). Both ring realizations support refocusing, with different preferred fluxes and winding correlations. The two protocols also differ in duration, frequency spacing, and the objective used to select and calibrate the pulse. This second example therefore demonstrates persistence of refocusing in another ring protocol, not an isolated circumference response or a matched performance comparison between the two sizes.

\FloatBarrier
\section{Discussion}

The fixed-pulse comparison and winding reconstruction identify coherent return as the source of the high fidelity. Constrained optimization exploits this interference. The two return geometries support refocusing at different phases while retaining the common-node separation.

Coherent recovery is already used in leakage-interference gates \cite{Rol2019LeakageInterference} and in nonadiabatic BIC-assisted transfer through neighboring scattering states \cite{Guo2026BICTransfer}. The distinction here is the remote control of propagation outside the transfer subspace while the transfer states, their energies, and the motion-induced coupling remain fixed. The flux can therefore tune return without changing the geometric coupling that launches the excursion. This acts separately from slowing the subspace motion or canceling the transitions it generates \cite{Bergmann1998,Vitanov2017,Berry2009}. It also differs from Aharonov--Bohm caging, where flux reorganizes propagation throughout the lattice \cite{Vidal1998,Longhi2014,Mukherjee2018}.

Common nodes provide a practical way to locate the control. For a transfer subspace preserved by $H_0(t)$, let $\Pi_{\mathcal N}$ project onto fixed sites where every column of $B(t)$ vanishes throughout the pulse. Since $\Pi_{\mathcal N}B(t)=0$, a perturbation $V_\chi(t)=\Pi_{\mathcal N}W_\chi(t)\Pi_{\mathcal N}$ with $W_\chi(t)$ Hermitian satisfies $V_\chi P=PV_\chi=0$. Couplings between these nodes can change propagation of an escaped excitation while leaving the compact states intact. Flux control also requires a closed path supporting a gauge-invariant loop phase; a static hopping phase on an open chain can be removed by rephasing the site basis. The node condition locates an allowed control, but useful refocusing still requires suitable relative phases of the returning amplitudes.

Subspace motion is essential to this response. A stationary projector admits a fixed frame with $D=0$. For $b(0)=0$, Eq.~(\ref{eq:b_projected}) then gives $b(t)=0$, and the encoded evolution has no dependence on $\chi$. Reducing $D$ and tuning complementary propagation are distinct control choices, coupled through the memory equation. In the single-excitation sector, the same propagation mechanism applies to coherent wave networks. Quantum transfer further requires recovery of the amplitudes and relative phases of every encoded superposition.

\section{Conclusion}

A remote flux can improve compact-state transfer by refocusing nonadiabatic excursions rather than preventing them. Common nodes allow the flux to reshape external propagation without changing the transfer states or their motion-induced coupling. In the ring, several winding-dependent matrix amplitudes interfere to suppress the endpoint error for every encoded input state. Fixed-pulse comparisons isolate this mechanism, and constrained optimization uses it to enhance worst-input fidelity. The design therefore preserves the instantaneous compact states while exploiting the coherent return of amplitude that leaves them during motion.

\section*{ACKNOWLEDGMENTS}
This work was supported by the National Natural Science Foundation of China under Grant Nos.~12675024 and 12275193.

\section*{DATA AVAILABILITY}

The data and custom code that support the findings of this article are available from the corresponding author upon reasonable request.

\appendix

\section{Projected dynamics and exact elimination of the complement}
\label{app:moving_projector}

\subsection{Projector identities}

Let $B(t)$ be an orthonormal frame for the moving encoded subspace, so that $B^\dagger B=I$ and $P=BB^\dagger$. It follows immediately that $P^2=P=P^\dagger$. Differentiating $P^2=P$ gives
\begin{equation}
	\dot P P+P\dot P=\dot P.
\end{equation}
Multiplication by $P$ from both sides yields
\begin{equation}
	P\dot P P=0.
	\label{app:eq:PdotP}
\end{equation}
With $Q=I-P$, one obtains
\begin{equation}
	\dot P=P\dot P Q+Q\dot P P.
	\label{app:eq:projector_motion}
\end{equation}
Thus the motion of a projector is purely off diagonal in the instantaneous decomposition $P\oplus Q$. Equation~(\ref{app:eq:projector_motion}) is basis independent. The matrix $D=C^\dagger\dot B$ is its $Q\leftarrow P$ representation in the chosen moving frames.

The conditions in Eq.~(\ref{eq:general_condition}) imply
\begin{align}
	B^\dagger H_\chi B&=B^\dagger H_0B,\\
	C^\dagger H_\chi B&=0,\\
	B^\dagger H_\chi C&=0,
\end{align}
for any orthonormal complementary frame $C(t)$. The control parameter therefore cannot enter the code Hamiltonian or the direct code--complement matrix elements. The only coupling between moving frames is geometric.

\subsection{Equations in the moving frame}

Insert $|\psi\rangle=Ba+Cb$ into $i|\dot\psi\rangle=H_\chi|\psi\rangle$. Projection with $B^\dagger$ gives
\begin{align}
	i\dot a+iB^\dagger\dot B\,a+iB^\dagger\dot C\,b
	=B^\dagger H_0B\,a.
\end{align}
Orthonormality of the combined frame implies
\begin{equation}
	B^\dagger\dot C=-\dot B^\dagger C=-D^\dagger,
	\qquad D=C^\dagger\dot B.
\end{equation}
Using Eq.~(\ref{eq:general_E}) therefore gives
\begin{equation}
	\dot a=-i\mathcal E a+D^\dagger b.
\end{equation}
Likewise, projection with $C^\dagger$ gives
\begin{equation}
	iDa+i\dot b+iC^\dagger\dot C\,b=C^\dagger H_\chi C\,b,
\end{equation}
and hence
\begin{equation}
	\dot b=-Da-i\mathcal H_{Q,\chi}b.
\end{equation}
The signs in Eqs.~(\ref{eq:a_projected}) and (\ref{eq:b_projected}) follow directly.

\subsection{Frame covariance}

The moving frames are not unique. Under time-dependent unitary rotations
\begin{equation}
	B'(t)=B(t)W_P(t),
	\qquad C'(t)=C(t)W_Q(t),
\end{equation}
the coefficients become $a'=W_P^\dagger a$ and $b'=W_Q^\dagger b$. The reduced operators transform as
\begin{align}
	\mathcal E'&=W_P^\dagger\mathcal EW_P-iW_P^\dagger\dot W_P,\\
	\mathcal H'_{Q,\chi}&=W_Q^\dagger\mathcal H_{Q,\chi}W_Q-iW_Q^\dagger\dot W_Q,\\
	D'&=W_Q^\dagger D W_P.
\end{align}
The complementary propagator and memory kernel obey
\begin{align}
	U'_{Q,\chi}(t,s)&=W_Q^\dagger(t)U_{Q,\chi}(t,s)W_Q(s),\\
	K'_\chi(t,s)&=W_P^\dagger(t)K_\chi(t,s)W_P(s).
\end{align}
Consequently, the projected equations and their physical predictions are frame covariant. A parallel frame is a convenience for the ring realization, not an assumption of the general theory.

\subsection{Exact elimination of the complement}

The propagator in Eq.~(\ref{eq:UQ_definition}) satisfies
\begin{equation}
	\partial_tU_{Q,\chi}(t,s)=-i\mathcal H_{Q,\chi}(t)U_{Q,\chi}(t,s).
\end{equation}
The composition law of the propagator gives its derivative with respect to the initial time,
\[
\partial_s U_{Q,\chi}(t,s)
=iU_{Q,\chi}(t,s)\mathcal H_{Q,\chi}(s).
\]
At fixed final time $t$, the complementary equation therefore implies
\[
\frac{d}{ds}[U_{Q,\chi}(t,s)b(s)]
=-U_{Q,\chi}(t,s)D(s)a(s).
\]
Integrating from $s=0$ to $s=t$ and using $U_{Q,\chi}(t,t)=I_Q$ gives
\begin{equation}
	b(t)=U_{Q,\chi}(t,0)b(0)
	-\int_0^tU_{Q,\chi}(t,s)D(s)a(s)\,ds.
\end{equation}
For $b(0)=0$, substituting this expression into $\dot a=-i\mathcal E a+D^\dagger b$ gives
\begin{align}
	\dot a(t)={}&-i\mathcal E(t)a(t)\nonumber\\
	&-\int_0^tD^\dagger(t)U_{Q,\chi}(t,s)D(s)a(s)\,ds,
\end{align}
which is Eq.~(\ref{eq:memory_equation}). The negative sign follows from the negative source in the complementary equation. This exact relation is time nonlocal and retains the complete complementary propagator.

For a periodic parameter, the coefficient
\begin{equation}
	U_{Q,\ell}(t,s)=\frac{1}{2\pi}\int_0^{2\pi}
	e^{-i\ell\chi}U_{Q,\chi}(t,s)\,d\chi
\end{equation}
gives Eqs.~(\ref{eq:UQ_fourier}) and (\ref{eq:kernel_fourier}) by Fourier inversion. Since $D(t)$ is parameter independent, projection acts coefficient by coefficient.

\section{Exact construction of the compact transfer states}
\label{app:compact}

\subsection{Single-frequency ansatz and short-arc recurrence}

For one channel $\alpha$, write
\begin{equation}
	|\Psi_\alpha\rangle=u_s|s,\alpha\rangle+u_r|r,\alpha\rangle
	+\sum_{n=0}^{N_c-1}\psi_n|n\rangle.
	\label{app:eq:ansatz}
\end{equation}
We seek an eigenstate of energy $\Omega_\alpha$ whose photonic component is confined to the short arc. The nodal and support conditions are
\begin{equation}
	\psi_0=\psi_d=0,
	\qquad \psi_n=0\quad(d<n<N_c).
	\label{app:eq:nodal_support}
\end{equation}
For $1\le n\le d-1$, the discrete Schr\"odinger equation is
\begin{equation}
	\Omega_\alpha\psi_n=-J(\psi_{n-1}+\psi_{n+1}).
	\label{app:eq:recurrence}
\end{equation}
The ansatz $\psi_n=\mathcal A_\alpha\sin(k_\alpha n)$ and the identity
\begin{equation}
	\sin[k_\alpha(n-1)]+\sin[k_\alpha(n+1)]
	=2\cos k_\alpha\sin(k_\alpha n)
\end{equation}
give $\Omega_\alpha=-2J\cos k_\alpha$. The receiver node requires
\begin{equation}
	k_\alpha=\frac{m_\alpha\pi}{d},
	\qquad 1\le m_\alpha\le d-1.
	\label{app:eq:standing_condition}
\end{equation}

\subsection{Endpoint cancellation and endpoint-regular form}

At $E=\Omega_\alpha$, the emitter equations are satisfied because $\psi_0=\psi_d=0$. The cavity equation at the sender reduces to

\begin{equation}
	g_su_s=J\mathcal A_\alpha\sin k_\alpha.
	\label{app:eq:sender_cancel}
\end{equation}

At the receiver,
\begin{equation}
	g_ru_r=J\psi_{d-1}.
	\label{app:eq:receiver_cancel_raw}
\end{equation}
The standing-wave condition implies
\begin{align}
	\sin[k_\alpha(d-1)]
	&=\sin(m_\alpha\pi-k_\alpha)\nonumber\\
	&=(-1)^{m_\alpha+1}\sin k_\alpha
	=\eta_\alpha\sin k_\alpha,
	\label{app:eq:parity_identity}
\end{align}
where $\eta_\alpha=(-1)^{m_\alpha+1}$. Thus

\begin{equation}
	g_ru_r=J\mathcal A_\alpha\eta_\alpha\sin k_\alpha.
\end{equation}

Choosing $\mathcal A_\alpha=g_sg_r$ avoids division by a coupling that vanishes at a pulse endpoint and gives
\begin{equation}
	u_s=Jg_r\sin k_\alpha,
	\qquad u_r=\eta_\alpha Jg_s\sin k_\alpha.
\end{equation}
This polynomial representation is precisely the unnormalized vector in Eq.~(\ref{eq:compact_state}).

\subsection{Normalization, orthogonality, and endpoint mapping}

The discrete sine functions satisfy
\begin{equation}
	\sum_{n=1}^{d-1}\sin\left(\frac{m\pi n}{d}\right)
	\sin\left(\frac{m'\pi n}{d}\right)=\frac d2\delta_{mm'}.
	\label{app:eq:sine_orthogonality}
\end{equation}
The squared norm is therefore
\begin{align}
	\langle\widetilde B_\alpha|\widetilde B_\alpha\rangle
	={}&J^2\sin^2k_\alpha(g_s^2+g_r^2)\nonumber\\
	&+g_s^2g_r^2\sum_{n=1}^{d-1}\sin^2(k_\alpha n)\nonumber\\
	={}&J^2\sin^2k_\alpha(g_s^2+g_r^2)+\frac d2g_s^2g_r^2,
\end{align}
which gives Eq.~(\ref{eq:normalization}). For $\alpha\ne\beta$, the atomic components are orthogonal because the emitters are distinguishable, and the photonic overlap vanishes by Eq.~(\ref{app:eq:sine_orthogonality}). Hence $\langle B_\alpha|B_\beta\rangle=\delta_{\alpha\beta}$.

We take $g_s,g_r\ge0$, require $g_s^2+g_r^2>0$ at all times, and impose $g_s(0)=g_r(T)=0$ and $g_r(0)=g_s(T)=g_{\max}>0$. Since $1\le m_\alpha\le d-1$ implies $\sin k_\alpha>0$, the normalized endpoint limits are
\begin{equation}
	|B_\alpha(0)\rangle=|s,\alpha\rangle,
	\qquad
	|B_\alpha(T)\rangle=\eta_\alpha|r,\alpha\rangle.
	\label{app:eq:endpoint_mapping}
\end{equation}
The compact photonic support is restricted to the interior of the short arc, so any remote bond $n_\phi\leftrightarrow n_\phi+1$ on the long arc obeys
\begin{equation}
	\langle n_\phi|B_\alpha(t)\rangle=
	\langle n_\phi+1|B_\alpha(t)\rangle=0.
\end{equation}
The phase-bearing bond is supported only on these sites and therefore satisfies Eq.~(\ref{eq:two_sided}).

\subsection{Parallel frame for the explicit basis}

Every $|B_\alpha(t)\rangle$ can be chosen real. Differentiating its normalization gives $\langle B_\alpha|\dot B_\alpha\rangle=0$. For $\alpha\ne\beta$, time differentiation changes only scalar coefficients; the atomic-label overlaps remain zero and the photonic overlap vanishes by discrete sine orthogonality. Therefore
\begin{equation}
	B^\dagger(t)\dot B(t)=0.
\end{equation}
A parallel complementary frame exists by solving
\begin{equation}
	\dot U_C=-C_0^\dagger\dot C_0U_C,
	\qquad C=C_0U_C,
\end{equation}
for any differentiable starting frame $C_0(t)$. Since $C_0^\dagger\dot C_0$ is anti-Hermitian, $U_C$ stays unitary and $C^\dagger\dot C=0$.

\section{Cyclic cover and winding interpretation}
\label{app:cover}

\subsection{Finite cyclic cover and infinite-cover limit}

A finite construction fixes the normalization and Fourier signs before the infinite cover is taken. Cut the bond $n_\phi\leftrightarrow n_\phi+1$ and make $N_\phi$ copies of the opened graph, labeled by $q=0,\ldots,N_\phi-1$ with $q+N_\phi\equiv q$. The crossing term on the cover is

\begin{equation}
	\widetilde V^{(N_\phi)}=-J\sum_{q=0}^{N_\phi-1}
	\left(|n_\phi,q+1\rangle\langle n_\phi+1,q|+\mathrm{H.c.}\right).
	\label{app:eq:finite_cover_crossing}
\end{equation}

For $\phi_j=2\pi j/N_\phi$, define the normalized Bloch states

\begin{equation}
	|\mu;\phi_j\rangle=\frac{1}{\sqrt{N_\phi}}
	\sum_{q=0}^{N_\phi-1}e^{-iq\phi_j}|\mu,q\rangle.
	\label{app:eq:finite_bloch}
\end{equation}

With the orientation in Eq.~(\ref{app:eq:finite_cover_crossing}),
\begin{equation}
	\langle n_\phi;\phi_j|
	\widetilde V^{(N_\phi)}
	|n_\phi+1;\phi_j\rangle
	=-Je^{i\phi_j},
	\label{app:eq:finite_cover_phase_check}
\end{equation}
which reproduces the phase convention of Eq.~(\ref{eq:ring_control}). These Bloch states block diagonalize the cover Hamiltonian. If $\widetilde U^{(N_\phi)}_{\ell0}(T,0)$ is the propagator block from cell $0$ to cell $\ell$ modulo $N_\phi$, Fourier inversion gives

\begin{align}
	U(T,0;\phi_j)&=\sum_{\ell=0}^{N_\phi-1}e^{i\ell\phi_j}
	\widetilde U^{(N_\phi)}_{\ell0}(T,0),
	\label{app:eq:finite_cover_propagator}\\
	\widetilde U^{(N_\phi)}_{\ell0}(T,0)&=\frac{1}{N_\phi}\sum_{j=0}^{N_\phi-1}
	e^{-i\ell\phi_j}U(T,0;\phi_j).
	\label{app:eq:finite_cover_inverse}
\end{align}

The positive sign in Eq.~(\ref{app:eq:finite_cover_propagator}) follows directly from the $e^{-iq\phi_j}$ convention. Letting $N_\phi\to\infty$ removes the identification of cell indices modulo $N_\phi$ and yields the infinite cyclic cover,

\begin{equation}
	U(T,0;\phi)=\sum_{\ell\in\mathbb Z}e^{i\ell\phi}
	\widetilde U_{\ell0}(T,0).
	\label{app:eq:cover_propagator}
\end{equation}

Projection onto the endpoint registers yields $A_\ell=\mathsf R^\dagger\widetilde U_{\ell0}(T,0)\mathsf S$ and hence Eq.~(\ref{eq:A_fourier}). Reversing the cut orientation reverses both the cell displacement and the Fourier sign, leaving the physical reconstruction unchanged. Because $P(t)$, $Q(t)$, and the chosen moving frames are independent of $\phi$, the same decomposition applies coefficientwise to $U_Q(t,s;\phi)$ and to $K_\ell(t,s)=D^\dagger(t)U_{Q,\ell}(t,s)D(s)$.

\subsection{Fixed cut and gauge convention}

Throughout the $N_c=41$ winding analysis, the cut is the bond between sites 29 and 30, its orientation is the one used in Eq.~(\ref{app:eq:finite_cover_crossing}), and the physical sender and receiver bases are fixed. Within this convention, each $A_\ell$ is an unambiguous matrix block of the cover propagator. Moving the cut or relocating the Peierls phase by a gauge transformation can relabel or rephase individual sector matrices. The coherently reconstructed transfer map, expressed in the correspondingly transformed endpoint bases, yields the same physical amplitudes and logical fidelities. We therefore assign physical significance to the full reconstruction and to interference between sectors within the stated convention, not to an isolated $A_\ell$ independently of that convention.

\section{Worst-input fidelity and numerical range}
\label{app:fidelity}

The sender input and receiver target are $|\psi_s(c)\rangle=\mathsf S c$ and $|\psi_r(c)\rangle=\mathsf R c$. The sans-serif symbols $\mathsf S$ and $\mathsf R$ are reserved for the endpoint embeddings, whereas $\mathcal R(\phi)$ denotes the return-error matrix in Eq.~(\ref{eq:return_error}). After propagation and receiver calibration, the target overlap is $c^\dagger ZAc$. Minimizing its squared modulus gives Eq.~(\ref{eq:Flogic}). Population outside the receiver subspace has zero target overlap and is penalized without postselection.

For any square matrix $M$, introduce
\begin{equation}
	H_\vartheta(M)=\frac{e^{-i\vartheta}M+e^{i\vartheta}M^\dagger}{2}.
\end{equation}
The supporting line of the numerical range normal to $e^{i\vartheta}$ satisfies
\begin{equation}
	\min_{z\in W(M)}\operatorname{Re}(e^{-i\vartheta}z)
	=\lambda_{\min}[H_\vartheta(M)].
\end{equation}
Convex separation from the origin gives
\begin{equation}
	d[0,W(M)]=\max\left\{0,\max_\vartheta
	\lambda_{\min}[H_\vartheta(M)]\right\}.
\end{equation}
If $0\in W(M)$, the distance is zero. Otherwise, the supporting line normal to the closest point gives the largest positive separation from the origin, which proves the preceding expression. Setting $M=ZA$ proves Eq.~(\ref{eq:numerical_range}). The smallest receiver population is $P_R^{\min}=\lambda_{\min}(A^\dagger A)$. For unitary $Z$ and normalized $c$,

\begin{equation}
	|c^\dagger ZAc|\le\|Ac\|,
\end{equation}
by the Cauchy--Schwarz inequality. Let $c_{\min}$ be a normalized eigenvector of $A^\dagger A$ with eigenvalue $P_R^{\min}$. Evaluating the minimum in Eq.~(\ref{eq:Flogic}) on this particular vector gives

\begin{align}
	F_{\rm logic}(Z,A)
	&\le |c_{\min}^\dagger ZAc_{\min}|^2\nonumber\\
	&\le c_{\min}^\dagger A^\dagger A c_{\min}
	=P_R^{\min}.
	\label{app:eq:receiver_population_bound}
\end{align}

Unlike $F_{\rm logic}$, $P_R^{\min}$ does not diagnose relative phase errors or channel mixing.

The most general diagonal correction used here is
\begin{equation}
	Z=e^{i\zeta_0}\operatorname{diag}(1,e^{i\zeta_2},e^{i\zeta_3})A_{\rm id}^{-1}.
\end{equation}
The common phase drops out of $F_{\rm logic}$, leaving two relative phases. A full $U(3)$ receiver correction is excluded. In the winding analysis, $Z_\star$ and $\theta_\star$ are selected once at $\phi_\star$ and then frozen.

To fix the otherwise irrelevant common phase, let $M_\star=Z_\star A(\phi_\star)$ and define

\begin{align}
	f(\theta)
	&=\left\|I_3-e^{-i\theta}M_\star\right\|_F^2\nonumber\\
	&=3+\|M_\star\|_F^2
	-2\operatorname{Re}\!\left[e^{-i\theta}\operatorname{Tr}M_\star\right].
	\label{app:eq:global_phase_objective}
\end{align}

The real part is maximized when $e^{-i\theta}\operatorname{Tr}M_\star$ is real and nonnegative, which minimizes $f(\theta)$. If $\operatorname{Tr}M_\star\ne0$, the minimizing phase is therefore

\begin{equation}
	\theta_\star=\arg\operatorname{Tr}M_\star,
	\label{app:eq:global_phase_minimizer}
\end{equation}
which is Eq.~(\ref{eq:theta_star}). If $\operatorname{Tr}M_\star=0$, the objective is independent of $\theta$ and the common phase may be chosen arbitrarily.

If $e^{-i\theta_\star}Z_\star A=I-\mathcal R$, then
\begin{equation}
	|c^\dagger Z_\star Ac|
	=|1-c^\dagger \mathcal Rc|
	\ge1-\|\mathcal R\|_2\ge1-\|\mathcal R\|_F,
\end{equation}
so
\begin{equation}
	F_{\rm logic}\ge\max\{0,1-\|\mathcal R\|_F\}^2.
\end{equation}
Here $\|\cdot\|_2$ denotes the spectral norm and $\|\cdot\|_F$ the Frobenius norm.
All reported values of $F_{\rm logic}$ are evaluated directly from the numerical range rather than from this sufficient bound.

\raggedbottom
\section{Bounded control family}
\label{app:control}

The endpoint pulses are generated from ten real coefficients through
\begin{equation}
	u(t)=1+\sum_{m=1}^{5}\left[a_m\cos(m\Delta\Omega t)
	+b_m\sin(m\Delta\Omega t)\right],
	\label{app:eq:shape_control}
\end{equation}
\begin{equation}
	f(t)=\int_0^t\sin^2\left(\frac{\pi\tau}{T}\right)u(\tau)\,d\tau,
	\label{app:eq:fcontrol}
\end{equation}
and
\begin{equation}
	g_s(t)=g_{\max}\frac{f(t)}{f(T)},
	\qquad g_r(t)=g_s(T-t).
	\label{app:eq:controls}
\end{equation}

\begin{figure}[t]
	\centering
	\includegraphics[width=0.91\columnwidth]{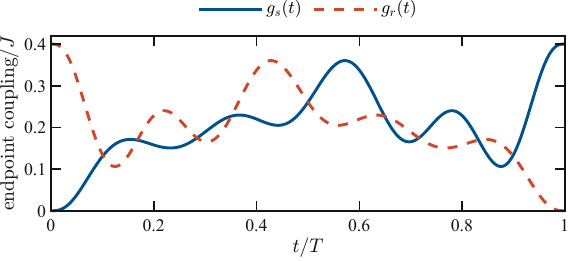}
	\caption{Endpoint couplings for the $N_c=41$, $JT=75$ refocusing protocol. The pulse satisfies $g_s(0)=g_r(T)=0$, $g_s(T)=g_r(0)=0.4J$, the common amplitude bound, the nonvanishing condition in Eq.~(\ref{app:eq:admissibility}), and the zero-slope endpoint conditions in Eq.~(\ref{app:eq:endpoint_slopes}).}
	\label{fig:control_pulses}
\end{figure}

The endpoint couplings of the refocusing protocol used in the fixed-pulse analysis are plotted in Fig.~\ref{fig:control_pulses}.

For the $N_c=41$, $JT=75$ comparisons, the common parameters and bounds are
\begin{equation}
	g_{\max}=0.4J,
	\qquad \Delta\Omega=0.06J,
	\qquad |a_m|,|b_m|\le3.
\end{equation}
The $N_c=43$, $JT=105$ example instead uses $\Delta\Omega=2\pi/T$ with the same five harmonics and bounds. For $N_c=43$, the function $\dot f(t)=\sin^2(\pi t/T)u(t)$ has a $T$-periodic extension with highest harmonic frequency $12\pi/T<0.4J$.

Let $\Omega_e=2\pi/T$ and $\omega_m=m\Delta\Omega$. Since $\sin^2(\pi t/T)=[1-\cos(\Omega_e t)]/2$, the constant term integrates to
\begin{equation}
	I_0(t)=\frac t2-\frac{\sin(\Omega_e t)}{2\Omega_e}.
\end{equation}
The cosine and sine terms give
\begin{align}
	I_m^{(c)}(t)={}&\frac{\sin(\omega_m t)}{2\omega_m}
	-\frac{\sin[(\omega_m+\Omega_e)t]}{4(\omega_m+\Omega_e)}
	\nonumber\\
	&-\frac{\sin[(\omega_m-\Omega_e)t]}{4(\omega_m-\Omega_e)},\\
	I_m^{(s)}(t)={}&\frac{1-\cos(\omega_m t)}{2\omega_m}
	-\frac{1-\cos[(\omega_m+\Omega_e)t]}{4(\omega_m+\Omega_e)}
	\nonumber\\
	&-\frac{1-\cos[(\omega_m-\Omega_e)t]}{4(\omega_m-\Omega_e)}.
\end{align}
The limiting expression is used if a denominator vanishes, and
\begin{equation}
	f(t)=I_0(t)+\sum_{m=1}^{5}\left[a_mI_m^{(c)}(t)+b_mI_m^{(s)}(t)\right].
\end{equation}
Equation~(\ref{app:eq:controls}) imposes the desired endpoint values. Admissible coefficient vectors are required to satisfy

\begin{equation}
	\begin{aligned}
		f(T)&>0,\\
		0&\le g_s(t),g_r(t)\le g_{\max}\quad (0\le t\le T),\\
		g_s^2(t)+g_r^2(t)&>0.
	\end{aligned}
	\label{app:eq:admissibility}
\end{equation}
The construction also gives the endpoint values
\begin{equation}
	\begin{aligned}
		g_s(0)&=0, & g_s(T)&=g_{\max},\\
		g_r(0)&=g_{\max}, & g_r(T)&=0.
	\end{aligned}
	\label{app:eq:endpoint_values}
\end{equation}
Since $\dot f(t)=\sin^2(\pi t/T)u(t)$, the endpoint slopes also vanish:
\begin{equation}
	\dot g_s(0)=\dot g_s(T)=\dot g_r(0)=\dot g_r(T)=0.
	\label{app:eq:endpoint_slopes}
\end{equation}

The largest frequency appearing in $\dot g_s(t)$ and $\dot g_r(t)$ is
\begin{equation}
	\omega_{\max}=5\Delta\Omega+\frac{2\pi}{T}\simeq0.384J
\end{equation}
at $JT=75$.

Thus $\omega_{\max}$ bounds the frequencies in the chosen synthesis, although a time-limited pulse is not strictly band limited.

\section{Winding identities and flux noise}
\label{app:winding}

Using Eq.~(\ref{eq:return_error}),
\begin{align}
	\frac{1}{2\pi}\int_0^{2\pi}\|\mathcal R(\phi)\|_F^2\,d\phi
	={}&\frac{1}{2\pi}\int_0^{2\pi}\sum_{\ell,m}e^{i(\ell-m)\phi}
	\nonumber\\
	&\times\operatorname{Tr}(\mathcal R_m^\dagger \mathcal R_\ell)\,d\phi
	\nonumber\\
	={}&\sum_\ell\|\mathcal R_\ell\|_F^2=P_{\rm inc}.
\end{align}
Grouping the off-diagonal terms by $q=\ell-m>0$ gives
\begin{equation}
	P_{\rm cross}(\phi)=2\sum_{q\ge1}
	\operatorname{Re}[e^{iq\phi}\Gamma_q],
\end{equation}
and hence Eq.~(\ref{eq:error_harmonics}). A symmetric sector set defines
\begin{equation}
	\mathcal R^{(L)}(\phi_\star)=\sum_{|\ell|\le L}e^{i\ell\phi_\star}\mathcal R_\ell.
\end{equation}
The corresponding phase-aligned calibrated transfer is

\begin{equation}
	I-\mathcal R^{(L)}(\phi_\star)
	=e^{-i\theta_\star}Z_\star A^{(L)}(\phi_\star).
\end{equation}

Because the logical fidelity is a nonlinear function of the coherent matrix sum, no additive-probability interpretation applies to the sector reconstruction.

For quasistatic flux noise, let the applied phase be $\phi_\star+\delta$, where $\delta$ is a zero-mean Gaussian variable of variance $\sigma_\phi^2$. Since $\langle e^{iq\delta}\rangle=e^{-q^2\sigma_\phi^2/2}$,

\begin{equation}
	\left\langle\|\mathcal R\|_F^2\right\rangle
	=P_{\rm inc}+2\sum_{q\ge1}e^{-q^2\sigma_\phi^2/2}
	\operatorname{Re}\left[e^{iq\phi_\star}\Gamma_q\right].
\end{equation}

Flux noise leaves the diagonal sector power unchanged and damps correlations according to their winding difference. This result is exact for the stated quasistatic Gaussian phase model and fixed calibration; it is not an average of the nonlinear worst-input fidelity.

%

\end{document}